\documentclass{egpubl_pg_short}
\usepackage{pg2020s}
 
\WsPaper           % uncomment for final version of EG Workshop contribution
\usepackage[T1]{fontenc}
\usepackage{dfadobe}  

\usepackage{cite}  % comment out for biblatex with backend=biber
\BibtexOrBiblatex
\electronicVersion
\PrintedOrElectronic
\ifpdf \usepackage[pdftex]{graphicx} \pdfcompresslevel=9
\else \usepackage[dvips]{graphicx} \fi

\usepackage{egweblnk} 
\usepackage{lipsum}
\usepackage{lineno}
\usepackage{setspace}
\usepackage{algorithm}
\usepackage{algorithmic}
\usepackage{xcolor}
\usepackage{tikz}
\usepackage{subcaption}

\definecolor{FIXMECOLOR}{rgb}{1,0,0}

\definecolor{TMPCOLOR}{rgb}{0,0,1}

\definecolor{CMTCOLOR}{rgb}{0.8,0,0.5}

\newcommand{\Vector}[1]{\mbox{\bf #1}}

\newcommand{\norm}[1]{\left\Vert {#1} \right\Vert}

\let\mat=\mathbf%
\let\set= \mathcal%

\title[Feature-aware Sparse Mesh Representation]
{A Robust Feature-aware Sparse Mesh Representation\vspace{-1em}}

\author[Fuentes et al.]
{\parbox{\textwidth}{\centering 
        Lizeth J. {Fuentes Perez}$^{1}$\orcid{0000-0003-1096-2871},
        Luciano A. {Romero Calla}$^{1}$\orcid{0000-0002-6186-4027},
        Anselmo A. Montenegro$^{2}$\orcid{0000-0001-9342-8732},
        Claudio Mura$^{1}$\orcid{0000-0002-6017-557X} and
        Renato Pajarola$^{1}$\orcid{0000-0002-6724-526X}
        }
        \\
{\parbox{\textwidth}{\centering 
        $^1$Visualization and MultiMedia Lab, University of Zurich, Switzerland \qquad
        $^2$Institute of Computing, Federal Fluminense University, Brazil
       }
}
}

\begin{document}

% uncomment for using teaser
% \teaser{
%  \includegraphics[width=\linewidth]{eg_new}
%  \centering
%   \caption{New EG Logo}
% \label{fig:teaser}
%}

\maketitle
%-------------------------------------------------------------------------
\begin{abstract}

The sparse representation of signals defined on Euclidean domains has been successfully applied in signal processing. Bringing the power of sparse representations to non-regular domains is still a challenge, but promising approaches have started emerging recently. 
% although promising works have been recently proposed. 
%
In this paper, we investigate the problem of sparsely representing discrete surfaces and propose a new representation that is capable of providing tools for solving different geometry processing problems. The sparse discrete surface representation is obtained by combining innovative approaches into an integrated method. First, to deal with irregular mesh domains, we devised a new way to subdivide discrete meshes into a set of patches using a feature-aware seed sampling. Second, we achieve good surface approximation with over-fitting control by combining the power of a continuous global dictionary representation with a modified Orthogonal Marching Pursuit. The discrete surface approximation results produced were able to preserve the shape features while being robust to over-fitting. 
% Hence, we affirm that the method is quite promising for surface re-sampling. Moreover, the level of mesh compression obtained was also satisfactory considering that we used a simpler dictionary build on top of fixed orthogonal basis functions.
Our results show that the method is quite promising for applications like surface re-sampling and mesh compression.
   
%-------------------------------------------------------------------------
%  ACM CCS 1998
%  (see https://www.acm.org/publications/computing-classification-system/1998)
% \begin{classification} % according to https://www.acm.org/publications/computing-classification-system/1998
% \CCScat{Computer Graphics}{I.3.3}{Picture/Image Generation}{Line and curve generation}
% \end{classification}
%-------------------------------------------------------------------------
%  ACM CCS 2012 see https://www.acm.org/publications/class-2012)
%The tool at \url{http://dl.acm.org/ccs.cfm} can be used to generate
% CCS codes.
%Example:


\end{abstract}  

% \linenumbers

%----------------------------------------------------------------------------------------------
\section{Introduction}

Triangle meshes are the \emph{de facto} standard for representing surface models in computer graphics, thanks to their ability to provide piecewise-linear approximations of continuous surfaces using only a finite number of primitives. 
Despite their advantages, the specific nature of triangle meshes makes them ill-suited for many common problems in polygonal mesh processing, which require specialized representations and the use of particular methods connected to such representations. Examples of such problems include the extraction of multiple levels of detail~\cite{H:98}, mesh compression and decimation~\cite{KCS:98}, re-sampling~\cite{VC:04} and the computation of discrete differential information~\cite{CP:03, DHM:03, HKP:11}.

% A variety of approaches were proposed to tackle these problems. For compact representation, decimation \cite{KCS:98} and re-sampling methods \cite{VC:04} were proposed. For multiresolution and level of detail, specific data-structures and control strategies\cite{H:98} were devised. For discrete differential information computation, the Osculating Jets method\cite{CP:03} and methods that work directly on the discrete representation \cite{DHM:03, HKP:11} were proposed.
%
Using a combination of these approaches in the same workflow is common in practice, but is not practical when multiple different operations need to be performed on the same surface model, as this requires storing multiple representations or converting from one to another that supports the desired operation. This creates a strong case for a more flexible representation of 3D surface models that combines the simplicity of triangle meshes with the ability to describe the underlying surface in a smooth manner. Since 3D shapes are typically encoded as triangle meshes, the target representation should be easy to construct from a high-quality input mesh, allowing to discard much of its redundancy in the construction process. In this context, we investigate the use of sparse representations for meshes and propose a representation that is a promising way to solve different geometry processing problems in a unified manner.

Sparse representations have been used extensively for regular domain signal processing including time-series, image and video processing. The main issues regarding their use for non-regular domains (especially meshes) lie in how to represent the concept of patches, how to deal with sampling differences at each patch and how to adapt or devise new ways to compute sparse codes based on a chosen basis function or learned dictionary.

In this work, we propose a solution for sparse mesh representation based on the combination of a feature-aware surface subdivision and on the use of a continuous, global, dictionary-based sparse representation sensitive to the sampling of the original mesh. In particular, we treat the surface underlying a mesh as a collection of height-fields defined over local tangent domains and express each of them as a sparse combination of 2D continuous functions. We show in the experiments that our method can approximate the surface without being impacted by over-fitting problems, which favors later surface re-sampling. We also obtained quite promising results in the context of mesh compression, even without using a learned dictionary. 
Our work represents a significant first step towards the development of a sparsity-based unified framework for efficient geometry processing. 

\section{Related work}\label{sec:related_work}
The use of sparse representations has been extensively investigated in computer vision and signal processing, mostly for signals defined in Euclidean domains. 
Recently, researchers have started to extend existing approaches to irregular domains like 3D meshes and point clouds, leading to new, specialized approaches to define the sparse codes, to structure the input data and to cast the problem as a constrained optimization. We discuss here only the most relevant works, referring the reader to recent surveys~\cite{XWZYDCL:15, LOMTB:18} for a more comprehensive review.

Digne et al.~\cite{DCV:14} have considered the problem of compressing highly detailed points clouds by leveraging self-similarity: they express local neighborhoods in a sparse manner, using a dictionary learned from local patches in a resampled version of the input cloud. We follow a similar general approach, but propose a different patch extraction technique that is well-suited to mesh domains and that does not require costly co-variance computations; moreover, our patch definition allows for an adaptive coverage of the input surface that accounts for local detail.

In a more recent work~\cite{DVC:18}, Digne and colleagues use a statistical method based on \textit{local probe fields} to define local patches based on shape self-similarity, using them to create a dictionary used for sparse representation. The main drawback of this statistical approach is that important features that do not occur frequently are not accounted for in their dictionary; to avoid this problem, we define patches using a feature-aware sampling process.

The closest work to ours is the surface approximation method  by Xu et al.~\cite{XWYDCL:16}, which is based on coupled optimization of sparsity and parameterization transformation. They show that this technique allows to faithfully represent sharp features, overcoming overfitting artifacts by the composition of smooth monomials and non-smooth domain optimization. We deal with a similar problem, but base our parameterizaton on the partitioning of the surface into patches that represent
2D functions defined over local tangent plane of the surface. In this sense, our work uses the patch definition proposed by Gu\'erin et al.~\cite{GDGP:16} for sparse terrain representation, but extends it to work with surfaces of arbitrary topology.
Like Xu and colleagues, we also deal with overfitting, but do so using a simple yet effective adaptation of the Orthogonal Matching Pursuit (OMP) algorithm, using the Nyquist sampling theorem to correctly account for the sampling rate of the patches.

%----------------------------------------------------------------------------------------------
\section{Method}\label{sec:patches_description}

%----------------------------------------------------------------------------------------------
\def\simages#1#2#3#4
{
	\begin{figure}[!htb]
		\centering
		\foreach \imagepath/\imagename/\imagecaption in #2
		{
			\subcaptionbox{\label{\imagepath/\imagename} \em \imagecaption}
			{\includegraphics[width=#1\linewidth]{\imagepath/\imagename}}
		}
		\caption{#4}\label{#3}
	\end{figure}
}

\def\images#1#2#3#4
{
	\begin{figure*}[!htb]
		\centering
		\foreach \imagepath/\imagename/\imagecaption in #2
		{
			\subcaptionbox{\label{\imagepath/\imagename} \em \imagecaption}
			{\includegraphics[width=#1\linewidth]{\imagepath/\imagename}}
		}
		\caption{#4}\label{#3}
	\end{figure*}
}
%---------------------------------------------

%The methodology of the proposed method is depicted in Figure~\ref{fig:method}.

%----------------------------------------------------------------------------------------------
\def\Patch{\set{P}}
\def\Vertices{\set{V}}
\def\Seeds{\set{S}}
\def\Faces{\set{F}}
\def\myalgorithm#1#2
{
\begin{algorithm}[H]
	\caption{#2}\label{#1}
	\input{#1}
\end{algorithm}
}

%----------------------------------------------------------------------------------------------

%\paragraph*{Mesh subdivision-based patches}\label{sec:patches_description}
\begin{comment}
\begin{figure}[H]
  \centering
  \includegraphics[width=0.45\textwidth]{figures/method.pdf}
  \caption{Methodology. Feature aware sampling (left), patch creation and normalization (middle). Sparse coding (right)}
  \label{fig:method}
\end{figure}
\end{comment}
\begin{comment}
In this section we describe the proposed method, which  is divided into three main steps: sampling, patch construction and sparse representation.
\end{comment}

Our method takes as input a triangular mesh $T = (\Vertices,\Faces)$, consisting of the vertex positions $\Vertices$ and indexed triangles $\Faces$. From the set of vertices $\Vertices$, we choose a subset of points $\Seeds$ that will be the center points of the patches. We define each center point of a patch as a \textit{seed point} $s_p$. A patch $\Patch$ is defined as a set of neighborhood points within radius $r_p$ and must satisfy the conditions of being the result of sampling a function over a 2D domain with disk topology. To ensure this, we devised a specialized feature-based sampling method.
Finally, we represent the patches in a sparse manner using an optimized version of Orthogonal Matching Pursuit (OMP).

\begin{comment}
\begin{equation}
    \Patch =\{ v \in \Vertices \hspace{0.2cm} \mid \hspace{0.2cm} \norm{v - s_p} \leq r_p\},\hspace{0.25cm} \Patch \subseteq \Vertices, \hspace{0.25cm}  s_p \in \Seeds \subseteq \Vertices
\end{equation}
\end{comment}

%-------------------------------------------------------------------------------------------
\subsection{Sampling}

Our feature-aware sampling takes as input a set of features of the mesh ranked according to an importance value. Our method generates this set of weighted feature points $\set{I}$ using the Harris3D algorithm~\cite{SB:11}, since it is robust to various defects on meshes; the set $\set{I}$, ordered by increasing importance of its points, serves as input for our sampling algorithm. The importance value corresponds to the Harris operator value; points with higher importance value are points with higher saliency.

\myalgorithm{algorithms/sampling}{Feature sampling algorithm.}
Alg.~\ref{algorithms/sampling} describes the main steps of our feature-based sampling. A certain degree of overlap between patches is allowed and controlled by $\sigma$. The radius $r_p$ defines the patch size and represents an important attribute, since it has a strong influence on the mesh reconstruction result. When the sizes of the patches are too large, the results tend to exhibit over-smoothing artifacts. Therefore, we introduce variable patch radii instead of a single, fixed one, as we want to maximize the area of patches in smooth regions and limit the patch size in non-smooth regions. This helps us maintain the fine details where needed. 
Note that, besides considering the sheer distance between vertices, we define a set of additional conditions in the patch construction process, as described in more detail in the following section.

%Note that for the patch construction we impose a set of conditions, detailed in the following section. These conditions ensure that even if the distance between points are satisfied, the candidate seed point may still not be chosen.

%\CMT{ok, this is good now, you see one has to be precise even in pseudo code to make sure everything is clear I agree}

%-------------------------------------------------------------------------------------------
\subsection{Patch construction}\label{sec:patch_cons}

Given a potential seed point $s$, we compute the neighboring vertices $\set{N}$ ordered by distance to $s$. Then, we calculate the neighboring triangles $\set{T}$, ordered by increasing distance value. 
%\CMT{this is a bit unclear: a) ``a'' set sounds like there are many options, but here there's just one way of defining the neighbors, i.e. as 1-ring neighbors (or not?); b) mention that the distances of the vertices are used to define the distances of the triangles, and shortly describe how.} In our experiments, we use the geodesic distance because it is piece-wise smooth over the surface.
%
The method for creating patches is described in Alg.~\ref{algorithms/create_patch} and uses the following conditions to define whether a point should be included in a patch: 

\renewcommand{\Vector}[1]{\mathrm{#1}}
\begin{itemize}
 \item \emph{Smooth variation condition}. The angle between the normals $\Vector{n}_{t_1}, \Vector{n}_{t_2}$ of two  adjacent triangles $t_1, t_2$ must satisfy $\langle \Vector{n}_{t_1}, \Vector{n}_{t_2} \rangle < \delta$, where $\delta$ is a tolerance angle error for smoothness and it should be at most $\frac{\pi}{2}$ because a larger value will lead to point distributions with abrupt changes.
 For our experiments we consider $\delta = \frac{\pi}{6}$.
 
 \item \emph{Discrete normal cone condition}. The angle between the normal of the central vertex $s$ and every other triangle in the patch must be less than  $\frac{\pi}{2}$. This condition guarantees that all points which belong to the patch can be described by an (injective) function.
 
 \item \emph{Area condition}. Very large patches are prone to introduce errors when approximated using a set of functions. We therefore limit the size of each patch to prevent having patches with very unbalanced sizes. To this purpose, we compute the ratio between the area $A_p$ of the patch  and the total area $A$ of the mesh, and the ratio between the projected area $A^{\mathrm{proj}}_p$ of the patch and $A_p$. In order for a point to be included in a patch, at least one of the two ratios must be below a critical threshold value: 
 $$ \frac{A_p}{A} \leq \rho\  \hspace{0.15cm}\vee\hspace{0.15cm} \frac {A^{\mathrm{proj}}_p}{A_p} \leq \eta \quad .$$

Intuitively, $\rho$ represents a percentage of the area and $\eta$ is the maximum distortion allowed between the area of the patch and the projected area; note that the projected area is computed over the plane defined by the normal of the patch $\Vector{n}_p$. These two sub-conditions compensate each other because while the second tends to create very unbalanced sizes, the first one regulates the sizes of the patches when possible. The first condition allows a patch to grow more when it is restricted by the second one. The second condition ensures that a flat patch region does not stop growing due to the area restriction of the first condition.
\end{itemize}

\myalgorithm{algorithms/create_patch}{ Patch construction algorithm.}

Having extracted patches that satisfy these conditions, we compute the principal curvature directions (maximal $\Vector{T}_1$, minimal $\Vector{T}_2$) and the normal vector $\Vector{N}$ of each patch; these define an orthogonal frame and its corresponding transformation matrix $R = (\Vector{T}_1, \Vector{T}_2,N)$. 
The matrix $R$ defines a local coordinate system, mapping local and global coordinates for each patch. 
We use this transformation to express the points of each patch in the local reference frame, then re-scale the patches to fit in a disk of radius $r_p = 1$. 

Given these normalized patches expressed in their local system, we regard each patch as a function that maps each patch point to its height over the patch domain. As described in Sec.~\ref{sec:sparse_representation}, we compute a sparse representation for each of these functions using a modified sparse coding technique. Once the sparse codes have been computed, we compute a separate approximation of each patch, locally. Since  patches are allowed to have overlaps, a given vertex of the mesh can be included in the approximation of several patches. To obtain a unique representation of each vertex, we include a final \emph{blending} step in our pipeline. This is done in two steps: first, we express all patches in the global reference system using the inverse of $R$; then, the final position of each vertex (expressed in the original coordinate space) is computed by averaging its different approximate representations defined by the patches in which it is contained. 

%----------------------------------------------------------------------------------------------
\subsection{Sparse representation}\label{sec:sparse_representation}
%----------------------------------------------------------------------------------------------
\def\SpCodes{\mat{\alpha}}
\def\Dict{\mat{D}}
\def\fnquist{f}
%----------------------------------------------------------------------------------------------

%----------------------------------------------------------------------------------------------

The main idea of sparse representation is that a signal $Y$ can be decomposed into a sparse linear combination of functions called \emph{atoms}, which are regarded as a base $\Dict$ called \emph{dictionary}.
Sparse coding is the operation that allows to obtain the coefficients $\hat{\SpCodes}$ (i.e., the \emph{encoding}) of the linear combination so that this representation is as sparse as possible~\cite{E:10}:

\vspace{-1em}
\begin{equation}\label{eq:sparsecod}
\displaystyle \hat{\SpCodes} = \mathrm{arg} \min_{\SpCodes} 
\norm{Y - \Dict \, \SpCodes} \hspace{0.5cm}
s. t. \hspace{0.5cm} \parallel \SpCodes
\parallel_0 \leq L \quad .
\end{equation}

The minimization ensures that the signal $Y$ is approximated with the least error possible, under the constraint that few atoms of the dictionary $\Dict$ are used.
The term $\norm{\SpCodes}_0$ corresponds to the maximum number of non-zero elements allowed in the sparse encoding, which in turn determines the maximum number of atoms that contribute to the representation. In this sense, it measures the sparsity of the representation, and is controlled by the regularization term $L$. As a result of this sparsity constraint on $\SpCodes$, only meaningful atoms are considered~\cite{E:10} in the representation.
Using the $l_0$ norm, finding a solution to Eq.~\ref{eq:sparsecod} corresponds to solving an NP-hard problem. However, relaxing the constraint defined by this choice of norm allows computing a good approximate solution to the problem. This can be achieved by using greedy algorithms such as Orthogonal Matching Pursuit (OMP), which chooses the best matching projections of multidimensional data onto a dictionary~\cite{CL:11}.

The choice of a proper dictionary has a significant impact on the performance of the sparse decomposition method. The atoms of the dictionary are either taken from a pre-defined set or learned to fit the input signal~\cite{AEB:06}. In any case, they are chosen so that they best adapt to \emph{patches} of the input domain.
Both patches and atoms can be easily defined when working with regular domains like images or volumes, but their definition becomes problematic for surfaces: the irregular nature of their domain easily leads to patches with a variable number of elements, and the atoms must be therefore defined in a way that allows capturing this irregularity. 

To solve this problem, we move from the work of Litany et al.~\cite{LRB:16}, who have proposed using learned \emph{continuous} functions as dictionary atoms and using a discretized version of such atoms to sparsely represent the input signal. In their approach, the dictionary is defined as $\displaystyle \Dict = \Phi \, A$, where $\Phi$ represents continuous basis functions and $A$ is the set of learned coefficients used to form the atoms as linear combinations of the basis functions.

We use a similar strategy based on continuous 2D functions to deal with the fact that our patches are composed of a variable number of points located at irregular positions.
Specifically, given a patch $\set{P}_y$, we consider the 2D projections of each point of $\set{P}_y$ on the tangent plane on which $\set{P}_y$ itself is defined; we then evaluate each continuous atom of the dictionary at such 2D projections, thus obtaining a discrete set of height values. The discrete sets of height values (one for each continuous atom) yield a discrete dictionary $\Dict_y = \Phi_y \, A$, to be used to represent patch $\set{P}_y$ in a sparse manner.

In practice, we represent $\set{P}_y$ as the vector $y$ of the height values of its points over the patch domain. Its sparse representation is obtained by combining the atoms of $\Dict_y$ in a sparse manner, according to the formula $\displaystyle y \approx \Dict_y \, A \, \SpCodes_y = \Phi_y \, A \, \SpCodes_y$. 
Differently from the formulation of Litany and colleagues, we do not learn the coefficients in $A$, and set instead $A$ to be the identity matrix $I$. In other words, our dictionary is represented by the cosine basis functions themselves.

\begin{comment}
\begin{figure}
  \includegraphics[width=0.5\textwidth]{figures/mse_comparison_plots.pdf}
  \caption{Example of function approximation of a patch with very low density. (a) $\fnquist = 0$, (b)  $\fnquist = 0.5$, (c)  $\fnquist = 1$, (d) $\fnquist = 0$ with the addition of extra samples.}
  \label{fig:comparison_mse_overfitting}
\end{figure}
\end{comment}

Besides the use of a continuous dictionary to adapt to the irregular mesh domains, another key aspect of our work is a modified formulation for the OMP algorithm that avoids overfitting when computing a sparse representation for the patches described in Sec.~\ref{sec:patches_description}. In our setting, the signal $Y$ to be approximated corresponds to the patch points. Inspired by the well-known Nyquist rate relative to signal sampling, we explicitly limit the frequency of the basis functions corresponding to the atoms that can be used in the sparse representation: the maximum frequency used is bound by the average spacing between samples in the patches computed on the mesh.

We set the maximum atom frequency by defining a lower bound on the corresponding wavelength $\lambda$. Let us denote by $\mu$ the median of the average distance between each point of each patch and its nearest neighbors (i.e., the 1-ring neighbors in the mesh); then, we impose the following constraint:   $\displaystyle 1 / 2 \cdot \lambda > \fnquist \cdot \mu \quad .
$

In the formula, $\fnquist \in [ 0,1 ] $ is a factor that allows a finer control of overfitting. The value of $\fnquist = 0$ is the default setting for the OMP algorithm, that is, no limit is set on the frequency of the atoms used; setting $\fnquist = 1$ restricts the choice of the atoms according to the standard Nyquist rate. To allow for some degree of overfitting, $\fnquist$ can be set to an intermediate value like $\fnquist= \frac{1}{2}$. In this way, the method avoids the greedy selection of atoms that oscillate unnaturally and that would excessively increase the total variation of the signal, leading to noise in resampling. We found this criterion to be particularly useful for patches in which the sampling density is low. 

\simages{0.3}{{
figures/patch0/$\fnquist = 0$,
figures/patch025/$\fnquist = 0.25$,
figures/patch05/$\fnquist = 0.5$
}}
{fig:patch_overfitting}{Example of function approximation of a patch. }

Fig.~\ref{fig:patch_overfitting} shows different examples of sparse patch representation using our Nyquist-based criterion; in particular, the plots show the 2D function corresponding to the sparse linear combination of the cosine basis functions in our dictionary. The result in~\ref{figures/patch0} is obtained with the conventional OMP formulation and is a clear case of overfitting, shown by the excessive oscillation of the function and caused by the overly high frequencies of the underlying atoms. We regularize these results by limiting the maximum frequency used through the parameter $\fnquist$ and by using additional points to compute the sparse approximation, sampling them on the surface defined by the mesh on the domain of the patch. As a result, we obtain a more natural function approximation that closely fits the patch points without  overfitting. In our experiments, we set this parameter to an intermediate value of $\fnquist \in [0,0.5]$ to prevent an over smoothed patch fitting. Higher values of $f$ could miss some frequencies necessary to approximate some parts of the patch when it is more complex. 

%%%%%%%%%%%%%%%%%%%%%%%%%%%%%%%%%%%%%%%%%%%%%%%%%%%%%%%%%%%%%%%%%%%%%%%%%%%%%%%%%%%%%

\begin{figure}
  \centering
  \includegraphics[width=0.87\columnwidth]{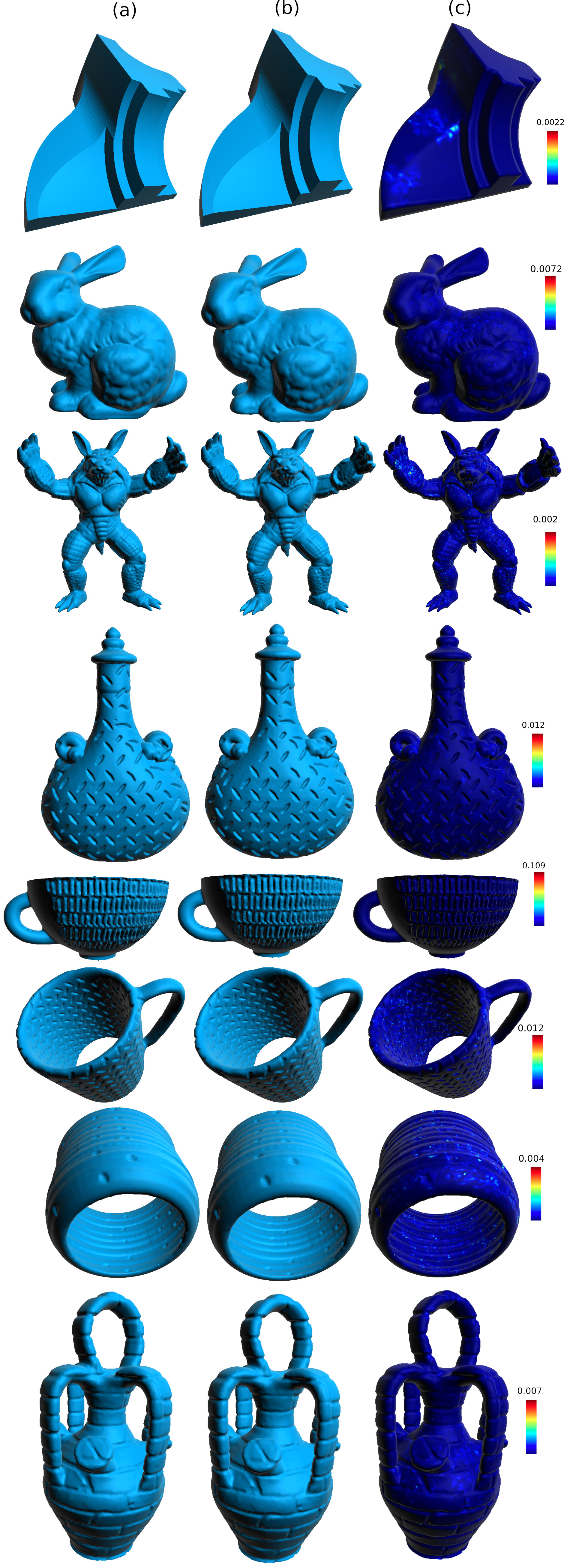}
  \caption{Visual comparison of approximation results. Original mesh $(a)$, Approximation result $(b)$. Colormap of error $(c)$. Models: \emph{fandisk}, \emph{bunny}, \emph{armadillo} and SHRECK models~\cite{MBGTWOBN:20}: \emph{109,165,170,202,72}.} 
  \label{fig:comparison_meshes}
\end{figure}

\simages{0.18}{{
figures/kit_sampling/,
figures/kit_original/,
figures/kit_rec/,
figures/kit_error/
}}
{fig:comparison_sampling}{Comparison between different sampling effect on approximation result. (a) Sampling, (b) original, (c) approximation results, (d) colormap of error. }

\section{Results}\label{sec:results}
%------------------------------------------------%

We tested our method on $11$ mesh models, which are listed in Tab.~\ref{tables/comparison} together with their vertex count $N_V$. These models are taken from common geometry processing benchmarks (including SHRECK ~\cite{MBGTWOBN:20}) and were selected to ensure that both smooth and sharp features are represented in our test suite. 

\def\mytable#1#2
{
\begin{table}[!htb]
	\caption{#2}\label{#1}
	\centering\input{#1}
\end{table}
}
%----------------------------------------------------%
\mytable{tables/comparison}{Approximation errors (RMSE): no. of vertices $N_V$, no. of patches $M$, percentage of area $\rho$, RMSE, Compression Ratio (C.R.)}

\begin{comment}
\begin{figure}
  \centering
  \includegraphics[width=0.3\textwidth]{figures/closeup.jpg}
  \caption{Close up comparison between the original Fandisk mesh (left) and the approximation result of our method (right).}
  \label{fig:comparison_closeup}
\end{figure}
\end{comment}

\begin{comment}
\simages{0.3}{{
figures/original/Original,
figures/xu/Xu et. al,
figures/ours/Ours
}}
{fig:comparison}{Comparison between the method proposed in \cite{XWYDCL:16} in \ref{figures/xu} and our approach \ref{figures/ours}. The original Fandisk mesh with $6475$ vertices \ref{figures/original}.}
\end{comment}

\begin{figure}
  \centering
  \includegraphics[width=0.45\textwidth]{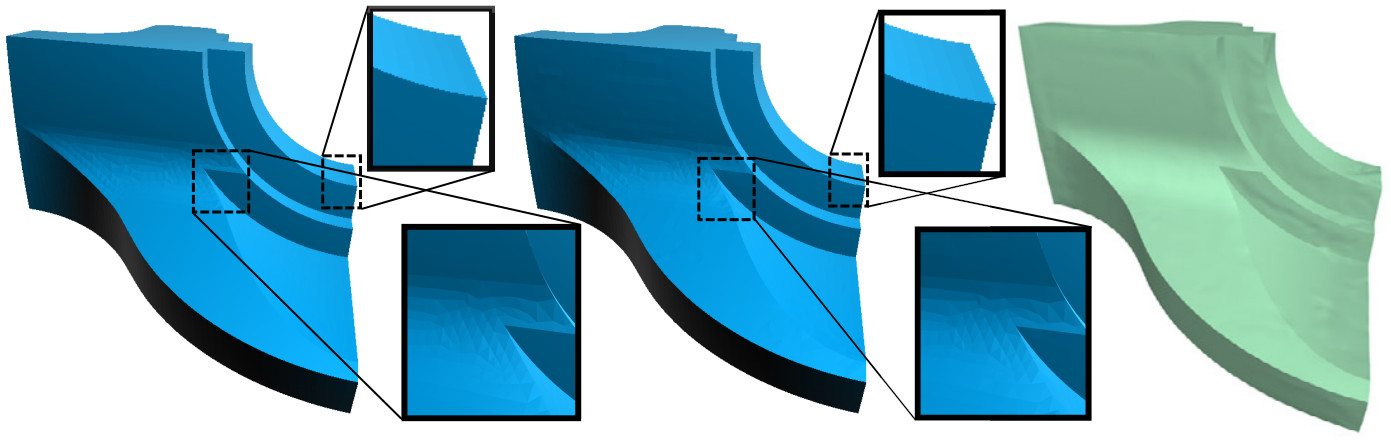}
  \caption{Comparison between the method proposed in \cite{XWYDCL:16} (right) and our approach (center). The original Fandisk mesh with $6475$ vertices (left).}
  \label{fig:comparison}
\end{figure}

We fixed the value of the parameters $\delta$, $\eta$ and $\fnquist$ to the values $\frac{\pi}{6}$, $1.01$ and $0.5$, respectively, because $\delta$ and $\eta$ do not depend on any global measure of the input mesh. The parameter $\eta$ depends solely on the local patch area and the projected area of the patch. The parameter $\delta$ depends on the local deviation angles of neighboring triangles in the patch.
We chose the value of $\rho$ empirically for each model (as shown in Tab.~\ref{tables/comparison}). This parameter changes for different meshes since it is influenced by the total area of each mesh. Also $\rho$ is not a fixed parameter, the user can increase it to obtain larger patches and decrease it to improve the accuracy of the approximation result, it serves to control the trade-off between error and compression ratio.
For all the experiments, we use a set of $144$ cosine basis functions as atoms of our (fixed) dictionary and used a minimum of $64$ samples to compute the sparse approximation.  

%The different values of $\rho$ for all the experiments shown in figures~\ref{fig:comparison_meshes},\ref{fig:comparison_sampling} and \ref{fig:comparison} are described in table~\ref{tables/comparison}. Also, we describe the number of vertices $n$, number of patches $M$, the MSE error and the compressed ratio (C.R).

For each model, we report in Tab.~\ref{tables/comparison} the number $M$ of patches extracted in the first stage of our method, the Root Mean Square Error (RMSE) incurred by approximating the input mesh with our output representation and the \emph{Compression Ratio} $CR$. The RMSE is obtained from the Euclidean distance between corresponding point pairs in both the original and reconstructed meshes and is given in distance units~\cite{XHFS:15}.
The value of $CR$ indicates the relative reduction in size obtained by representing the original model using our sparse model and is computed as follows:
\begin{equation}
\displaystyle CR = \frac{13 \cdot M + \norm{\SpCodes}_0}{3 \cdot N_V} \quad .
\end{equation}

In this formula, the numerator denotes the size of our sparse representation: besides storing the $\norm{\SpCodes}_0$ sparse codes for the mesh, we store for each of the $M$ patches $13$ values, corresponding to $9$ entries of the global-to-local alignment matrix $R$, to the $3$ coordinates of the seed points and to a single value for the radius $r_p$ of each patch.
The denominator indicates the size of the input representation; note that we define it only based on the number of vertices, omitting the storage required to store the face connectivity.

A visual overview of the approximation quality obtained with our algorithm is shown in Fig.~\ref{fig:comparison_meshes}. One can clearly see that the output approximation is virtually indistinguishable from the original. The high quality of our approximation is further confirmed by the colormap of the error: in this visualization, the error ranges from $0$ (blue) to the value $e_{max}$ (red), which is the maximum point-to-point Euclidean distance error.

\begin{figure}
  \centering
  \includegraphics[width=0.35\textwidth]{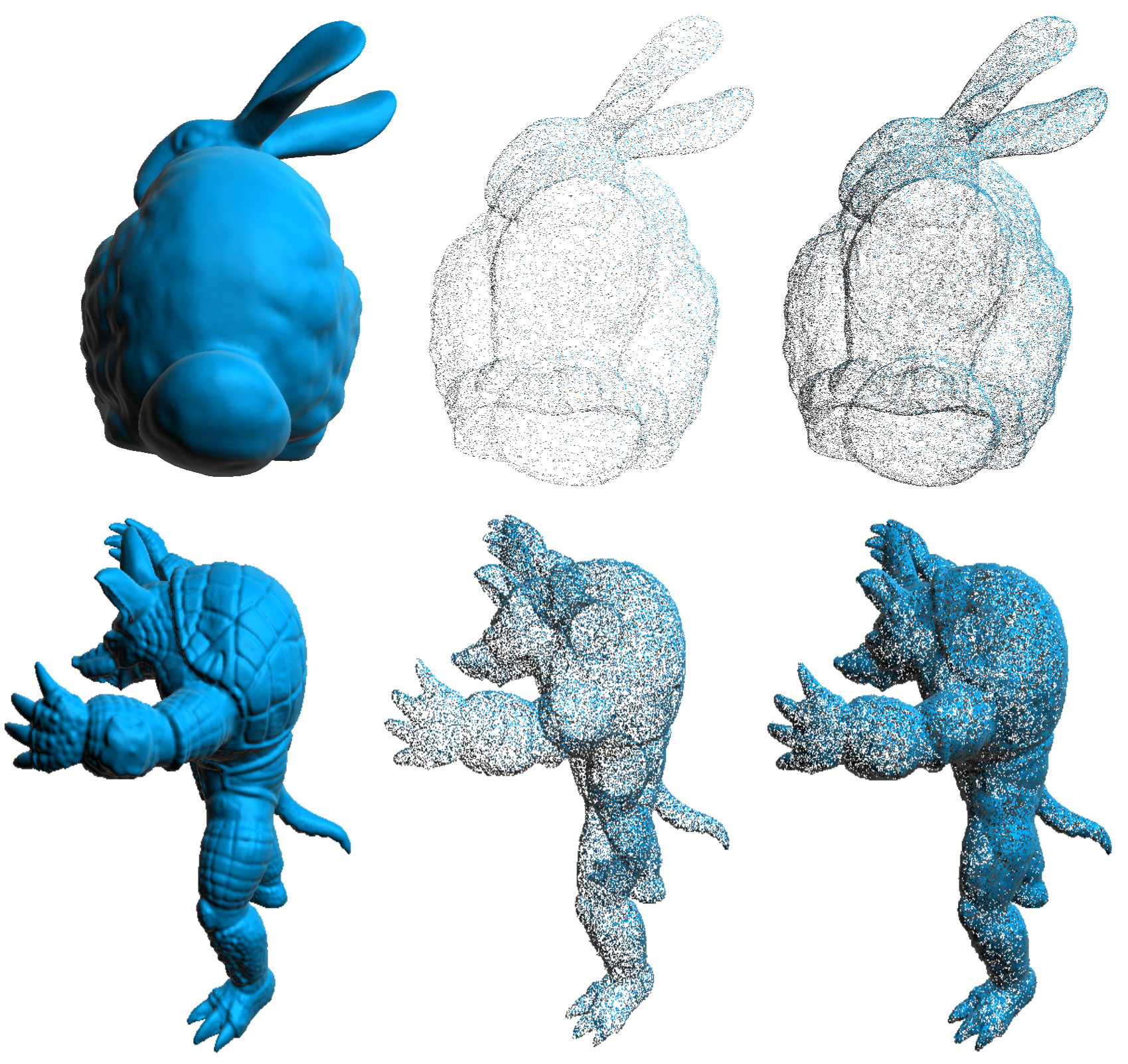}
  \caption{Visual comparison of resampling results. Original mesh (left), low density (middle), higher density (right).}
  \label{fig:comparison_resampling}
\end{figure}

%% Armadillo resampling was included in previous image for this version
\begin{comment}
\begin{figure}
  \centering
  \includegraphics[width=0.4\textwidth]{figures/arma_resampling.jpg}
  \caption{Visual comparison of resampling results in Armadillo mesh. Original mesh (left), low density (middle), higher density (right).}
  \label{fig:arma_resampling}
\end{figure}

\end{comment}

Our feature aware sampling can create regions of variable sizes that are adaptable to the mesh curvature. Patches with bigger and smaller sizes are created in smooth and high curvature regions, respectively. An example of the effect of varying the sampling density is depicted in Fig.~\ref{fig:comparison_sampling}, where the parameter $\rho$ is set to $ 0.001$ for the results in the top row and to $0.0002$ for those in the bottom row. As shown in the color-coded visualizations, decreasing the value of $\rho$ significantly reduces the approximation error, at the cost of an increased storage requirement for the representation: while only $M=860$ patches are required for $\rho=0.001$, this number increases to $1680$ in the case of $\rho=0.0002$.

%The number of vertices of the original Kitten shape in \ref{figures/kit_original} is $n=43544$. The sampling parameters are: $\delta = \frac{\pi}{6}$, $\eta = 1.01$ in both experiments. For the first row, the parameter $\rho = 0.001$ and the number of patches is $M=860$. For the second row, the parameter $\rho = 0.0002$ and the number of patches is $M=1680$.
%The colormap error is depicted in \ref{figures/kit_error}. The scale of the error is in the range $[0, 0.068]$, where $0.068$ is the maximum error. The mean square error (MSE), is $0.00041$ and $0.00029$ for the upper and lower row results, respectively. 
%The quantity of seeds depends on the parameters described in section~\ref{sec:patch_cons}. Note that the quality of the approximation result and the level of compression is linked to the number of seeds. 

Our technique allows for a faithful representation of sharp features, obtaining results comparable or superior to other state-of-the-art techniques. This can be evinced from the comparison between our approach and the method of Xu et al.~\cite{XWYDCL:16} shown in Fig.~\ref{fig:comparison}. %A close-up of our result is depicted in Figure~\ref{fig:comparison_closeup}. 
We can observe that our results exhibit significantly less noisy sharp features, while at the same time avoiding the need for a complex sparse representation formulation and for a costly parameterization to define the patches in pre-processing.

A natural application of our sparse representation is point-based mesh resampling. We create random samples in the 2D domain of each patch and recover the height values using our sparse representation. Examples of resampling at two different density levels are shown in Fig.~\ref{fig:comparison_resampling} (models \emph{bunny} and \emph{armadillo}). Note that, since we obtain a continuous, smooth representation of the original model, we are able to generate a point-based sampling at an arbitrary density level; moreover, we can compute a normal vector for each point in a straightforward manner from our functional representation, which allows for a clear, shaded visualization of the resampled point set without the need for a costly normal estimation step.

\section{Conclusions}\label{sec:conclusions}

In this paper we have proposed a practical method for approximating meshes via sparse representation. Our work is based on two key ingredients: a feature-aware strategy to organize an input mesh into a collection of patch-like local domains, over which the mesh can be defined as a height-field suitable for sparse representation; a sparse coding formulation to approximate each local height-field as a sparse combination of cosine basis functions, using a criterion derived from the well-known Nyquist sampling theorem to robustly control overfitting.

We evaluated our approach on a selection of mesh models that contain both smooth and sharp features, and used the resulting sparse approximation as a single representation to solve three common geometry processing problems: approximation, resampling and compression. We obtained good results even though we used a simple, fixed cosine basis as a dictionary.
In its current state, our method is not able to achieve competitive compression results on mesh models with a high degree of geometric detail. While the compression ratio could be improved by increasing the size of the patches, this would result in an increased complexity of the points distribution on the patches, making the cosine functions no longer able to approximate the surface in an adequate manner. 
As future work, we plan to improve the approximation by combining different types of basis functions and, more importantly, by learning a dictionary of new basis functions from the input data. 
\begin{comment}
We also intend to investigate the application of our approach to other problems such as mesh inpainting and computation of discrete differential information.
\end{comment}

\begin{spacing}{1.0}
\begin{footnotesize}
  \noindent\textbf{Acknowledgments.}
This work was partially supported by the EU MSCA-ITN project EVOCATION (grant agreement 813170) and by CAPES DS.
\end{footnotesize}
\end{spacing}\vspace{-1em}

% \begin{spacing}{0.8}
% \begin{scriptsize}
%   \noindent\textbf{Acknowledgments.} Acknowledgments go here
% \end{scriptsize}
% \end{spacing}

% bibtex
\bibliographystyle{eg-alpha-doi} 
\bibliography{refs}       

% biblatex with biber
% \printbibliography                

\end{document}